\documentclass[12pt]{iopart}

\usepackage{graphicx}
\usepackage{dcolumn}
\usepackage{bm}
\usepackage{braket}
\usepackage{enumerate}
\usepackage{color}
\usepackage{hyperref}
\usepackage{cite}
\begin{document}


\title[]{Distributed quantum information processing via single atom driving}

\author{Jing-Xin Liu, Jun-Yao Ye, Lei-Lei Yan, Shi-Lei Su$^*$}
\address{School of Physics and Engineering, Zhengzhou University, Zhengzhou 450001, China}
\ead{slsu@zzu.edu.cn}

\author{Mang Feng$^{\dag}$}
\address{State Key Laboratory of Magnetic Resonance and Atomic and Molecular Physics, Wuhan Institute of Physics and Mathematics, Chinese Academy of Sciences,
Wuhan 430071, China}
\ead{mangfeng@wipm.ac.cn}

\vspace{10pt}
\begin{indented}
\item[]\today
\end{indented}


\begin{abstract}
We propose an unconventional scheme for quantum entangled state distribution (QESD) and quantum state transfer~(QST) based on a fiber-cavity-atom system, in which three atoms are confined, respectively, in three bimodal cavities connected with each other by optical fibers. The key feature of the scheme is the virtual excitation of photons, which yields QESD and QST between the two atoms in the edge-cavities conditioned on one-step operation only on the atom in the middle cavity. No actual operation is performed on the two atoms in the edge cavities throughout the scheme. Robustness of the scheme over operational imperfection and dissipation is discussed and the results show that system fidelity is always at a high level. Finally, the experimental feasibility is justified using laboratory available values.
\end{abstract}


\maketitle


\section{Introduction}
Quantum entangled state distribution~(QESD), which aims to achieve quantum entanglement between distant nodes in quantum
network~\cite{PhysRevLett.78.3221,PhysRevLett.90.253601,PhysRevLett.91.067901}, plays critical role for quantum cryptography
implementation~\cite{PhysRevLett.67.661,PhysRevLett.81.5932}, quantum secret sharing~\cite{PhysRevLett.83.648}, quantum teleportation~\cite{PhysRevLett.70.1895},
and distributed quanutm computation~\cite{PhysRevA.59.4249}. So far, there have
been many schemes proposed for QESD using single-atoms~\cite{PhysRevLett.90.253601, PhysRevLett.83.5158, PhysRevLett.90.217902}, trapped
ions~\cite{PhysRevLett.91.067901,PhysRevLett.91.110405}, atomic ensembles~\cite{duan2001long}, nitrogen-vacancy centers~\cite{PhysRevLett.96.070504} as well as
cavity quantum
electrodynamics~\cite{kurpiers2018deterministic,PhysRevLett.91.177901,pyrkov2013entanglement,PhysRevLett.92.127902,PhysRevA.72.032333,hacker2016photon,tiecke2014nanophotonic,kimble2008quantum,PhysRevA.78.024302,PhysRevLett.92.013602,Shen:14,Jin:17}.
Besides, QESD in noisy channel~\cite{wang2009entanglement,sheng2010efficient}, even with long distance~\cite{yin2012quantum}, has also been well studied in photonic system. Fast QESD with atomic ensembles and fluorescent detection has also been studied~\cite{brask2010fast}. Quantum state
transfer~(QST)~\cite{PhysRevLett.78.3221,Mabuchi1372} intends to transmit quantum states (or quantum information) from one node to another in quantum network. The
mathematical form of the simplest QST between two nodes \emph{A} and \emph{B} can be expressed as
$|\psi\rangle_{A}|0\rangle_{B}\rightarrow|0\rangle_{A}|\psi\rangle_{B}$, where $|\psi\rangle$ is the transferred state. Like QESD, a lot of schemes have been
proposed for QST using atomic system~\cite{duan2001long, matsukevich2004quantum, PhysRevLett.90.253601, PhysRevLett.83.5158, PhysRevLett.90.217902}, trapped
ions~\cite{PhysRevLett.91.067901,PhysRevLett.91.110405}, spin chains~\cite{bose2003quantum, christandl2004perfect, yao2011robust},
superconducting~\cite{sillanpaa2007coherent,majer2007coupling,mei2018robust,xiang2017intracity}, and nitrogen-vacancy centers~\cite{PhysRevLett.96.070504}.
Besides, the dissipative dynamics has also been introduced into the QST working in circuit QED~\cite{WangandGertler} and Rydberg atom
systems~\cite{PhysRevA.99.032348}. Very recently, deterministic QESD and QST have been implemented experimentally in superconducting circuit
system~\cite{kurpiers2018deterministic} using microwave photons based on an all-microwave cavity-assisted Raman process.

\begin{figure}[htp]
\centering
\includegraphics[scale=0.8]{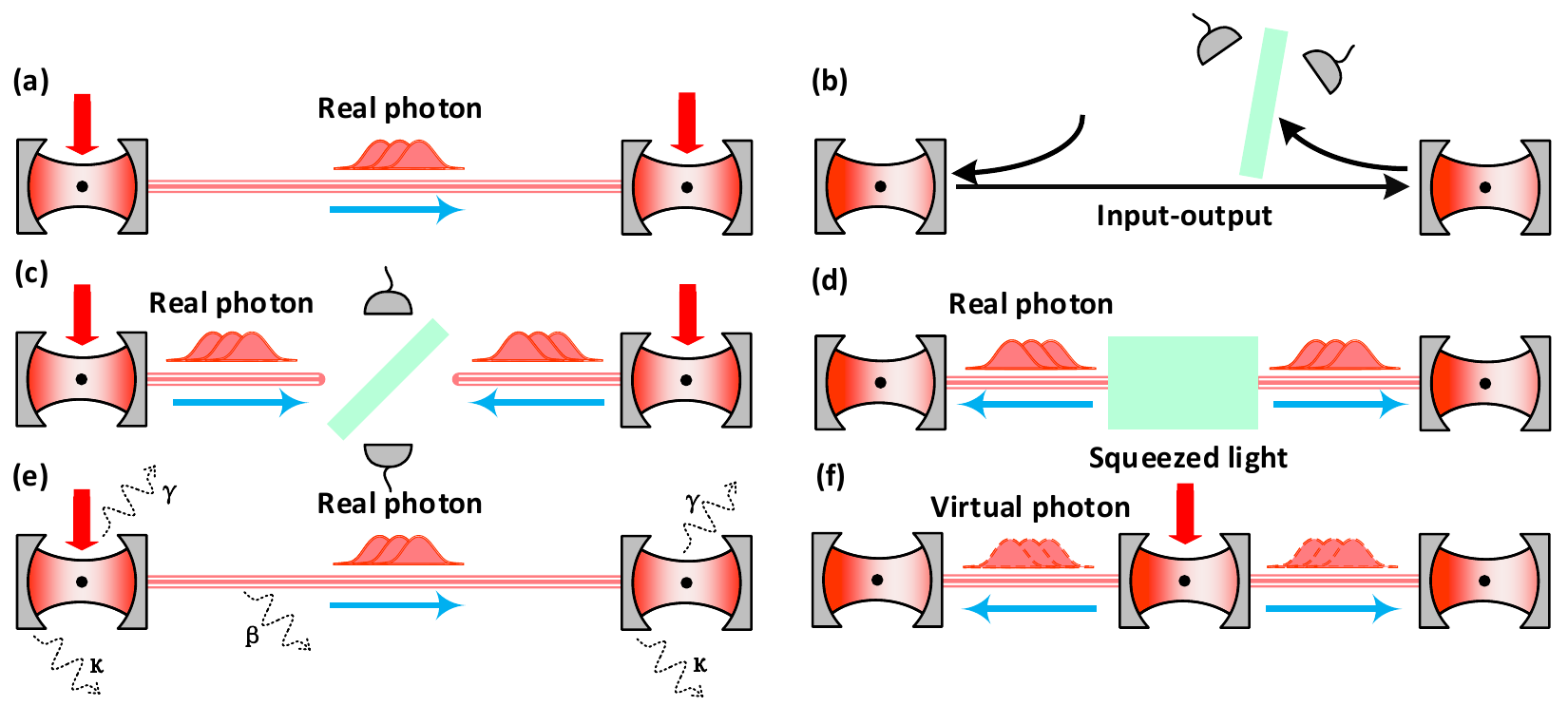}
\caption{Schematics for typical schemes to create long distance entanglement. (a) Input-output process of photons due to laser-driven atoms. By interacting with
the input photons of the laser,
the atom emits a photon to the cavity mode in the left node. Then the photon transmits to the right node through an optical fiber and interacts with the
laser-driven atom there to complete the QESD and
QST~\cite{PhysRevLett.78.3221,kurpiers2018deterministic,PhysRevLett.91.177901,kimble2008quantum,pyrkov2013entanglement} . (b) Similar to (a) but without laser
driving~\cite{PhysRevLett.92.127902,PhysRevA.72.032333,hacker2016photon,tiecke2014nanophotonic,kimble2008quantum,PhysRevA.78.024302}. (c) Interference of photons
from the atoms in different
nodes due to laser driving simultaneously~\cite{PhysRevLett.90.253601,duan2001long}. (d) Involvement of squeezed lights under dissipation. Two atoms trapped in
different nodes are driven by
squeezed lights simultaneously and then get entangled in steady states under dissipation~\cite{PhysRevLett.92.013602}. (e) Dissipative dynamics. One or both of the
laser-driven atoms in fiber-connected nodes distribute entanglement in a steady state~\cite{Shen:14,Jin:17}. (f) The present scheme.}
\label{fig:1}
\end{figure}

The QESD and QST schemes between two remote fiber-connected-cavities~(nodes) can be roughly categorized into following cases,  as sketched in Fig. \ref{fig:1}. For
cases (a, b, c), two separate nodes are operated one by one in sequence or simultaneously, and measurement of the output photons is required. For cases (d, e) with
dissipative dynamics involved, the QESD is achieved by a steady state due to competition between the drive and decoherence. But these two cases are not for QST,
which works based on unitary dynamics. The case (f) is a new scheme proposed in the present work, which, different from the previous QESD and QST
schemes, owns following favorable characteristics: (i) Two qubits employed for QESD and QST are not necessary under actual operations, but coupled/entangled due to
an auxiliary atom and virtually excited photons; (ii) The state of the auxiliary atom keeps invariant throughout the scheme, which makes the scheme robust to
decoherence. The paper is organized as follows. We first present an effective Hamiltonian for the atom-cavity-fiber model, based on which QESD and QST are implemented. Then we assess how well the scheme can be accomplished and how robust it is over the imperfection and dissipation. Experimental feasibility is justified based on laboratory available values. The result shows that the fidelity of system is more than 99.4\% by adjusting laser shape. Finally, we give a brief conclusion.

\section{The system and Hamiltonians}
\subsection{The basic model}

\begin{figure}
\centering
\includegraphics[scale=0.7]{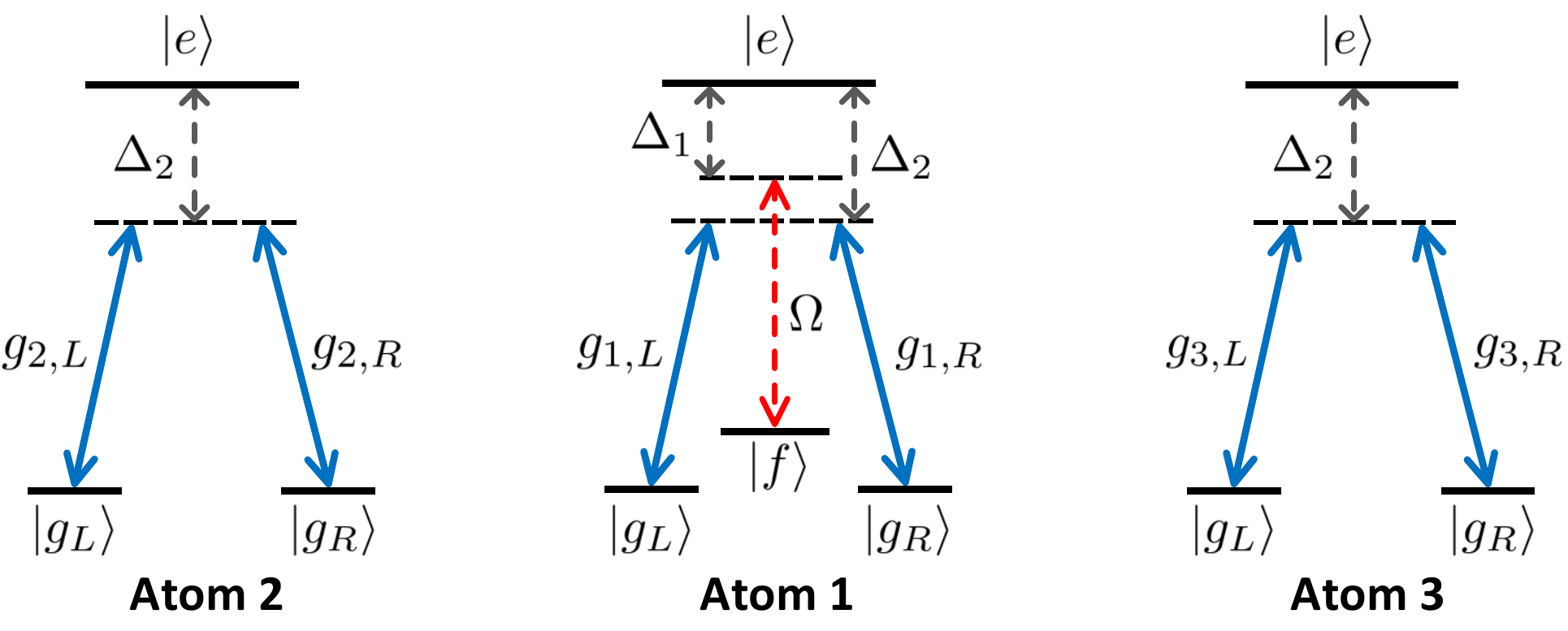}
\caption{Level scheme for Fig. 1(f), where the transitions $\ket{e}_n \leftrightarrow \ket{g_{L(R)}}_n$ ($n = 1,2,3$ denoting different atoms as labeled at the
bottom) are off-resonantly coupled to the left(right)-circularly polarized modes of the cavities. Detunings $\Delta_2$ exist in transitions $\ket{e}_n
\leftrightarrow \ket{g_{L(R)}}_n$ and the corresponding coupling constant is $g_{n,L(R)}$. Another laser is applied to off-resonantly drive the transition
$\ket{e}_1 \leftrightarrow \ket{f}_1$ in the atom $1$ with detuning $\Delta_1$ and Rabi frequency $\Omega$.}
\label{fig:2}
\end{figure}

Our scheme consists of three atoms confined, respectively, in three identical bimodal cavities connected by optical fibers. Each cavity, as detailed in Fig. \ref{fig:2}, contains a single
three-level atom interacting with the cavity by Jaynes-Cummings model \cite{jaynes1963jaynes} under rotating-wave
approximation \cite{shore1993bw}. In the interaction picture, the total Hamiltonian can be written, in units of $\hbar=1$, as
\begin{equation}
\eqalign{
\hat{H}_{I} &= \hat{H}_{CA}+\hat{H}_{LA} + \hat{H}_{CF}, \cr
\hat{H}_{CA} &=\sum_{k=1}^{3}\sum_{j=L,R}g_{k,j}\hat{a}_{k,j}\ket{e}_k\bra{g_j}e^{i\Delta_2 t}+{\rm H.c.}, \cr
\hat{H}_{LA} &= \Omega\ket{e}_1\bra{f}e^{i\Delta_1 t} + {\rm H.c.}, \cr
\hat{H}_{CF} &=\sum_{k=1}^{2}\sum_{j=L,R} \nu \hat{b}^{\dag}_{k,j}(\hat{a}_{1,j}+\hat{a}_{k+1,j}) + {\rm H.c.}, }
\label{eq:1}
\end{equation}
where $\hat{H}_{CA}$, $\hat{H}_{LA}$ and $\hat{H}_{CF}$ denote the cavity-atom interaction, laser-atom interaction and cavity-fiber interaction, respectively.
$\hat{a}_{k,j}~(k=1,2,3; j=L,R)$ is the annihilation operator of the $j$-circularly polarized mode of the cavity $k$; $\hat{b}_{k,j} ~(k=1,2)$ is the annihilation
operator of the $j$-circularly polarized mode of the optical fiber $k$;
$\nu$ is the coupling strength between the cavities and the fibers \cite{serafini2006distributed,pellizzari1997quantum}; $g_{k,j}$ is the coupling strength between
the atom $k~(k=1,2,3)$ and two circularly polarized modes of the cavity $k$. A laser field
is applied to the atom 1 with Rabi frequency $\Omega$. $\Delta_2$ and $\Delta_1$ are, respectively, detunings in the transitions $\ket{e}_n \leftrightarrow
\ket{g_{L(R)}}_n$ and $\ket{e}_1 \leftrightarrow\ket{f}_1$.

$H_{CF}$ is a working Hamiltonian for high fineness cavities under resonant operations over the time scale much longer than the fiber's round-trip time
\cite{serafini2006distributed,pellizzari1997quantum,van1999quantum} in the short
fiber limit. We assume the mode separation between neighboring fiber modes to be $\pi c/L$. This means that the number of the fiber modes coupling to the cavity
mode is of the order of $N = (2l\overline{\nu})/(2\pi c)$, where $\overline{\nu}$ is the cavity decay rate under the coupling with
the fibers and $c$ is the light speed in optical fibers. In this case, we set $N \le 1$ and the coupling of the cavity mode to an individual fiber mode can be
calculated approximately as $\sqrt{2\overline{\nu}\pi c/L}$. As such, there is only one resonant mode $\hat{b}_k$ of the fiber $k$ coupled between the adjacent
cavities.

\subsection{Effective Hamiltonian}

To have an insight into the significant nature of system, we first perform the following bosonic-mode transformation\cite{serafini2006distributed} for $\hat{H}_{I}$,
\begin{equation}
\eqalign{
\hat{c}_{\pm\sqrt{3},j}& = \frac{1}{2\sqrt{3}}(2 \hat{a}_{1,j} + \hat{a}_{2,j} + \hat{a}_{3,j} \pm \sqrt{3}\hat{b}_{1,j} \pm \sqrt{3}\hat{b}_{2,j} ), \cr
\hat{c}_{\pm,j}& = \frac{1}{2}( - \hat{a}_{2,j} + \hat{a}_{3,j} \mp \hat{b}_{1,j} \pm \hat{b}_{2,j}), \cr
\hat{c}_{0}& = \frac{1}{\sqrt{3}}(-\hat{a}_{1,j} + \hat{a}_{2,j} + \hat{a}_{3,j}).
}
\label{eq:2}
\end{equation}
which rewrites $\hat{H}_I$ as $\hat{H}^\prime_I = \hat{H}^\prime_{AC} + \hat{H}^\prime_{LA} + \hat{H}^\prime_{CF}$ with

\begin{equation}
\eqalign{
\hat{H}_{AC}^{\prime} &=\sum_{j=L,R}\frac{g_{1,j}}{\sqrt{3}}(\hat{c}_{+\sqrt{3},j}+\hat{c}_{-\sqrt{3},j}-\hat{c}_{0,j})\ket{e}_1\bra{g_j}e^{i\Delta_2 t} +
\sum_{k=2}^{3}\sum_{j=L,R}\frac{g_{k,j}}{2\sqrt{3}}
\cr
& \times [\hat{c}_{+\sqrt{3},j}+\hat{c}_{-\sqrt{3},j} + (-1)^{k-1}\sqrt{3}\hat{c}_{+,j} + (-1)^{k-1}\sqrt{3}\hat{c}_{-,j}+2\hat{c}_{0,j}] \cr
& \times \ket{e}_k\bra{g_j}e^{i\Delta_2 t}+{\rm H.c.}, \cr
\hat{H}_{LA}^{\prime} &= \Omega\ket{e}_1\bra{f}e^{i\Delta_1 t} + {\rm H.c.}, \cr
\hat{H}_{CF}^{\prime} &= \nu\sum_{j=L,R}(\sqrt{3}\hat{c}^{\dag}_{+\sqrt{3},j}\hat{c}_{+\sqrt{3},j} - \sqrt{3}\hat{c}^{\dag}_{-\sqrt{3},j}\hat{c}_{-\sqrt{3},j} + \hat{c}^{\dag}_{+,j}\hat{c}_{+,j} -
\hat{c}^{\dag}_{-,j}\hat{c}_{-,j}).
}
\label{eq:3}
\end{equation}
Then turning it into the interaction representation by performing the unitary operation $e^{-i(\hat{H}^{\prime}_{CF}-\Delta_1\sum_{k=1}^{3}\ket{e}_k\bra{e})t}$,  we obtain $\hat{H}_{I}^{\prime\prime} = \hat{H}_{AC}^{\prime\prime} + \hat{H}_{LA}^{\prime\prime}$ where
\begin{equation}
\eqalign{
\hat{H}_{AC}^{\prime\prime} &= \sum_{j=L,R}\frac{g_{1,j}}{\sqrt{3}}(\hat{c}_{+\sqrt{3},j}e^{i\delta_{+\sqrt{3}} t}+\hat{c}_{-\sqrt{3},j}e^{i\delta_{-\sqrt{3}} t}-\hat{c}_{0,j}e^{i\delta_0
t})\ket{e}_1\bra{g_j} +
 \cr
& \sum_{k=2}^{3}\sum_{j=L,R}\frac{g_{k,j}}{2\sqrt{3}}[\hat{c}_{+\sqrt{3},j}e^{i\delta_{+\sqrt{3}} t} +\hat{c}_{-\sqrt{3},j}e^{i\delta_{-\sqrt{3}} t} + (-1)^{k-1}\sqrt{3}\hat{c}_{+,j}e^{i\delta_+ t}\cr
&  + (-1)^{k-1}\sqrt{3}\hat{c}_{-,j}e^{i\delta_-t} + 2\hat{c}_{0,j}e^{i\delta_0 t}] \times \ket{e}_k\bra{g_j} + {\rm H.c.} \cr
}
\label{eq:4}
\end{equation}
and $\hat{H}_{LA}^{\prime\prime} = \Delta_1\sum_{j=1}^{3}\ket{e}_j\bra{e}+\Omega(\ket{e}_1\bra{f} +\ket{f}_1\bra{e})$ with the detuning satisfying $\delta_n=\Delta_2-\Delta_1+n\nu$ ($n=\pm\sqrt{3},\pm,0$).

By selecting suitable detuning $\delta_m = 0$, in the large detuning limit $\delta_n \gg g_{k,j}(n \neq m)$, Eq.~(\ref{eq:4}) can be reduced to a simple model that a bimodal cavity $c_{n,j}$ is coupled to an imaginary five-level atom system. All in all, the large detuning limit corresponds to $\nu \gg 1$. Choosing $\delta_{+}=0$ ($\delta_-=0$), under the rotating wave approximation only the terms containing $\hat{c}_{+,j}$ ($\hat{c}_{-,j}$) in Eq.~(\ref{eq:4}) is reserved. However these two modes $\hat{c}_{+,j}$ ($\hat{c}_{-,j}$) is decoupled with state $\ket{e}_1$ which implies that the atom 1 is out of interaction with the bimodal field. Similarly, if $\delta_{n}=0$ ($n=\pm\sqrt{3},0$), there are also only two modes $\hat{c}_{n,j}$. However, the state $\ket{e}_1$ is coupled to these two modes and a full coupling structure is obtained. In these situations, we write the effective Hamiltonians in a uniform form ($n=\pm\sqrt{3},0$),
\begin{equation}
\hat{H}_{eff} = \Delta_1\sum_{k=1}^{3}\ket{e}_k\bra{e}+[\Omega\ket{e}_1\bra{f}+\sum_{j=L,R}\bar{g}_j\hat{c}_{n,j}\ket{\vartheta}_j\bra{g_{j}} + {\rm H.c.}]
\label{eq:5}
\end{equation}
where we define the state $\ket{\vartheta}_j:=\sum_{k=1}^3\bar{g}_{k,j}\ket{e}_k/\bar{g}_j$ with the normalization coefficient $\bar{g}_j$, $\bar{g}_{1,j}=g_{1,j}/\sqrt{3}$ and $\bar{g}_{k,j}=g_{k,j}/2\sqrt{3}$ with $k=2,3$. (Some details can be seen in Appendix A.) Eq.~(\ref{eq:5}) is one of the main results in our model. To simplify the representation, we set $\bar{g}_{1,j} = g_{c1}$ and $\bar{g}_{k,j} = g_{c2}$ with $k=2,3$.


\section{Simplification in a subspace}

\begin{figure}
\centering
\includegraphics[scale=0.6]{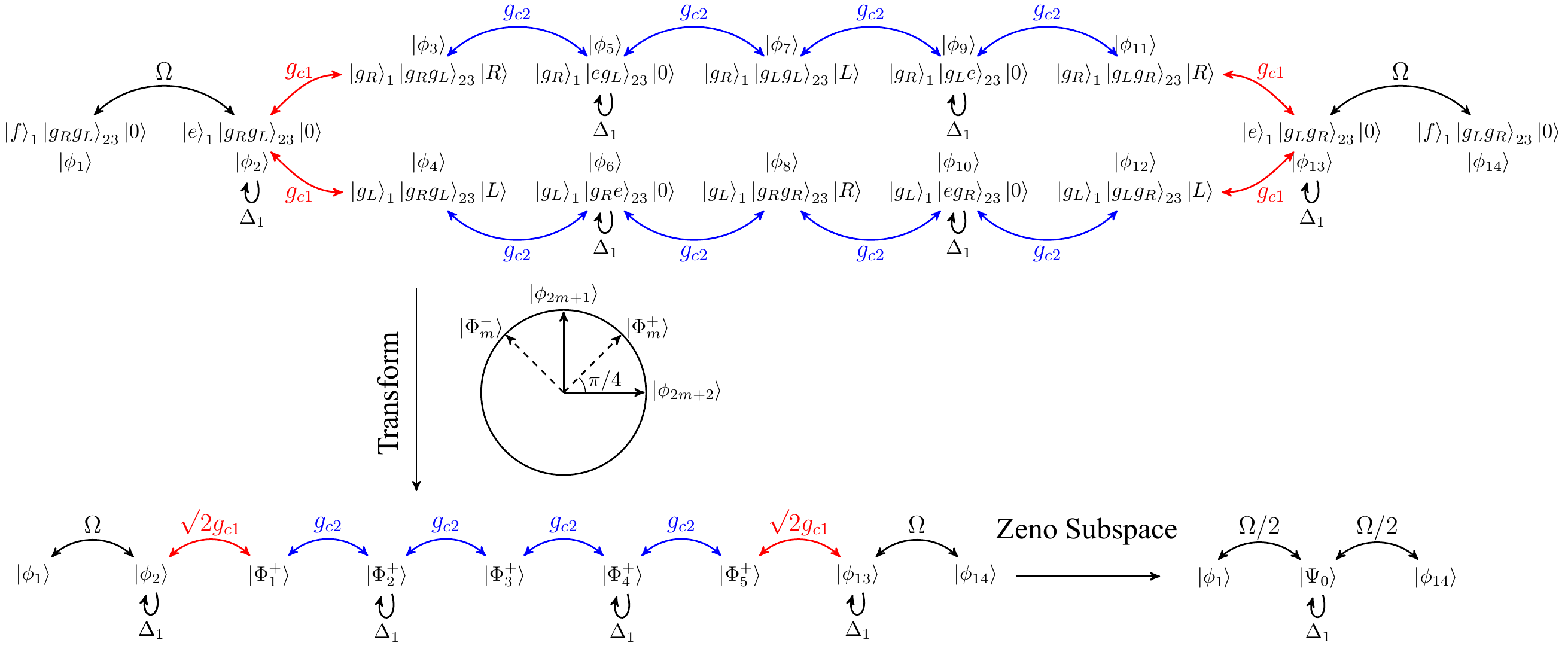}
\caption{Schematic for the simplification steps of Hamiltonians from Eq.~\ref{eq:6} to Eq.~\ref{eq:8}  and then to Eq.~\ref{eq:11} .}
\label{fig:4}
\end{figure}

For our purpose of achieving high-quality QESD and QST, we encode qubits in the ground states $\ket{g_Lg_R}_{23}$ and $\ket{g_Rg_L}_{23}$. To this end, we impose
the system to be initially in the state $\ket{\phi_0}=\ket{f}_1\ket{g_{R}g_{L}}_{23}\ket{000}_c\ket{00}_f$ denoting atoms 1, 2 and 3 in the states $\ket{f}$,
$\ket{g_{R}}$ and $\ket{g_{L}}$, respectively, and the fibers and three cavities in vacuum states. This initial state, after the bosonic-mode transformation in Eq.~(\ref{eq:2}), turns to be $\ket{\phi_0}\to\ket{\phi_1}=\ket{f}_1\ket{g_{R}g_{L}}_{23}\ket{0}$, as the initial state of the effective Hamiltonian described by
Eq.~(\ref{eq:5}). $\ket{0}$ in $\ket{\phi_1}$ in the vacuum state of bosonic mode in Eq.~(\ref{eq:5}). In order to describe $\hat{H}_{eff}$ in single-exciton space, we introduce new basis states $\{\ket{\phi_n}\}$, as given in Appendix, and
rewrite Eq.~(\ref{eq:5}) as below,
\begin{equation}
\eqalign{
\hat{\mathcal{H}}_{I} &= \hat{\mathcal{H}}_{g} + \hat{\mathcal{H}}_{\Omega} + \hat{\mathcal{H}}_{\Delta}, \cr
\hat{\mathcal{H}}_{g} &=  g_{c1}(\ket{\phi_2}\bra{\phi_{3}} + \ket{\phi_{13}}\bra{\phi_{11}} + \ket{\phi_2}\bra{\phi_{4}} + \ket{\phi_{13}}\bra{\phi_{12}}) \cr
& + g_{c2}(\ket{\phi_5}\bra{\phi_3} + \ket{\phi_8}\bra{\phi_{10}}) + g_{c2}(\ket{\phi_5}\bra{\phi_7} + \ket{\phi_{12}}\bra{\phi_{10}}) \cr
& + g_{c2}(\ket{\phi_6}\bra{\phi_8} + \ket{\phi_9}\bra{\phi_{11}}) + g_{c2}(\ket{\phi_6}\bra{\phi_4} + \ket{\phi_9}\bra{\phi_7}) + H.c. ,\cr
\hat{\mathcal{H}}_{\Omega} &= \Omega(\ket{\phi_2}\bra{\phi_1}+\ket{\phi_{13}}\bra{\phi_{14}}) + H.c. , \cr
\hat{\mathcal{H}}_{\Delta} &= \Delta_1(\ket{\phi_2}\bra{\phi_2} + \ket{\phi_5}\bra{\phi_5} + \ket{\phi_6}\bra{\phi_6} + \ket{\phi_9}\bra{\phi_9} +
\ket{\phi_{10}}\bra{\phi_{10}} + \ket{\phi_{13}}\bra{\phi_{13}}).
}
\label{eq:6}
\end{equation}
Eq.~(\ref{eq:6}) could be graphically understood in Fig. \ref{fig:4}. Because two paths exist in the coupling from $\ket{\phi_2}$ to $\ket{\phi_{13}}$, we further
consider a group of transformations,
\begin{equation}
\ket{\Phi_m^{\pm}}=\frac{1}{\sqrt{2}}(\ket{\phi_{2m+1}}\pm\ket{\phi_{2m+2}}), ~~(m=1,2,...,5)
\label{eq:7}
\end{equation}
and then by setting $\sqrt{2}g_{c1}=g_{c2}=g$, Eq.~(\ref{eq:6}) becomes
\begin{equation}
\eqalign{
\hat{\mathcal{H}}_{I} &= \hat{\mathcal{H}}_{g} + \hat{\mathcal{H}}_{\Omega} + \hat{\mathcal{H}}_{\Delta}, \cr
\hat{\mathcal{H}}_{g} &= g(\ket{\phi_2}\bra{\Phi_1^{+}}+\ket{\Phi_5^{+}}\bra{\phi_{13}}) + g\sum_{m=1}^{4}\sum_{n=\pm}\ket{\Phi_m^{n}}\bra{\Phi_{m+1}^{n}} + {\rm
H.c.},  \cr
\hat{\mathcal{H}}_{\Omega} &= \Omega(\ket{\phi_1}\bra{\phi_2}+\ket{\phi_{13}}\bra{\phi_{14}}) + {\rm H.c.}, \cr
\hat{\mathcal{H}}_{\Delta} &= \Delta_1[\ket{\phi_2}\bra{\phi_2}+\ket{\phi_{13}}\bra{\phi_{13}} + \sum_{n=\pm}(\ket{\Phi_2^{n}}\bra{\Phi_2^{n}} +
\ket{\Phi_4^{n}}\bra{\Phi_4^{n}})],
}
\label{eq:8}
\end{equation}
which implies that the system is effectively divided into two subspaces regarding $\{\ket{\Phi_m^+}\}$ and $\{\ket{\Phi_m^-}\}$ (See Fig. \ref{fig:4}). If the
system is initially prepared in $\ket{\phi_1}$ or $\ket{\phi_{14}}$, no state would evolve into the subspace regarding $\{\ket{\Phi_m^-}\}$. As such, in the
following treatment, we just consider the state evolution within a 9-dimensional Hilbert subspace spanned by
$\{\ket{\phi_1},\ket{\phi_2},\ket{\phi_{13}},\ket{\phi_{14}},\ket{\Phi_1^{+}},\ket{\Phi_2^{+}},\ket{\Phi_3^{+}},\ket{\Phi_4^{+}},\ket{\Phi_5^{+}}\}$.


\subsection{Zeno subspace}

\begin{table*}
\centering
\caption{\label{tab:1} Eigenvalues and eigenstates of $\hat{\mathcal{H}}_g$. Here we define $\xi_{\pm}=\sqrt{2 \pm \sqrt{2}}$ and $\eta_{\pm}=1 \pm \sqrt{2}$.}
\begin{tabular}{cc}
\br
Eigenvalues & Eigenstates \\ \mr
$\lambda=0$&$\ket{\Psi_0}=\frac{1}{2}(\ket{\phi_2} - \ket{\Phi_2^{+}} + \ket{\Phi_4^{+}} - \ket{\phi_{13}})$ \\
$\lambda_{1}^{\pm}=\pm\sqrt{2}g$&$\ket{\Psi_1^{\pm}} = \frac{1}{2\sqrt{2}}(\ket{\phi_2} \pm \sqrt{2}\ket{\Phi_1^{+}} + \ket{\Phi_2^{+}} - \ket{\Phi_4^{+}} \mp
\sqrt{2}\ket{\Phi_5^{+}} -
\ket{\phi_{13}})$ \\
$\lambda_{2}^{\pm}=\pm g \xi_{+}$&$\ket{\Psi_2^{\pm}} = \frac{\sqrt{2}}{4\xi_{+}}[\ket{\phi_2} \pm \xi_{+}(\ket{\Phi_1^{+}} + \sqrt{2}\ket{\Phi_2^{+}} +
\ket{\Phi_5^{+}} ) +
\eta_{+}(\ket{\Phi_2^{+}} + \ket{\Phi_4^{+}}) + \ket{\phi_{13}}]$ \\
$\lambda_{3}^{\pm}=\pm g \xi_{-}$&$\ket{\Psi_3^{\pm}} = \frac{\sqrt{2}}{4\xi_{-}}[\ket{\phi_2} \pm \xi_{-}(\ket{\Phi_1^{+}} - \sqrt{2}\ket{\Phi_2^{+}} +
\ket{\Phi_5^{+}} ) +
\eta_{-}(\ket{\Phi_2^{+}} + \ket{\Phi_4^{+}}) + \ket{\phi_{13}}]$ \\
\br
\end{tabular}

\end{table*}

In this section, we introduce Zeno conditions $\hat{\mathcal{H}}_g \gg \hat{\mathcal{H}}_\Omega$, which means $g \gg \Omega$, to simplify the dynamics of the system. After discarding the subspace regarding $\{\ket{\Phi_m^-}\}$, we rewrite the Hamiltonian Eq.\~(\ref{eq:8}) based on $\ket{\phi_1}$, $\ket{\phi_{14}}$ and the eigenstates of $\hat{\mathcal{H}}_g$(listed in Table. \ref{tab:1}),
\begin{equation}
\eqalign{
\hat{\mathcal{H}}^{\prime}_{I} &= \hat{\mathcal{H}}^{\prime}_{g}+\hat{\mathcal{H}}^{\prime}_{\Omega}+\hat{\mathcal{H}}^{\prime}_{\Delta}, \cr
\hat{\mathcal{H}}^{\prime}_{g} &= \sum_{m=1}^{3}\sum_{n=\pm}\lambda_{m}^{n}\ket{\Psi_m^{n}}\bra{\Psi_m^{n}}, \cr
\hat{\mathcal{H}}^{\prime}_{\Omega} &= \frac{\Omega}{2}[\ket{\Psi_0}+\frac{1}{\sqrt{2}}(\ket{\Psi_{1}^{+}} + \ket{\Psi_{1}^{-}})](\bra{\phi_1}-\bra{\phi_{14}}) \cr
& + \frac{\Omega}{4}(\ket{\phi_1}+\ket{\phi_{14}})[\xi_{-}(\bra{\Psi_2^{+}}+\bra{\Psi_2^-}) + \xi_{+}(\bra{\Psi_3^+} + \bra{\Psi_3^-})] + {\rm H.c.},  \cr
\hat{\mathcal{H}}^{\prime}_{\Delta} &= \Delta_1\ket{\Psi_0}\bra{\Psi_0}+\frac{\Delta_1}{2}\sum_{m=1}^{3}(\ket{\Psi_m^{+}} + \ket{\Psi_{m}^{-}})(\bra{\Psi_m^{+}} + \bra{\Psi_m^{-}}).
}
\label{eq:9}
\end{equation}
Eq.~(\ref{eq:9}) can be further simplified under a unitary transformation $e^{-i\hat{\mathcal{H}}^{\prime}_g t}$ and the condition of quantum Zeno
dynamics\cite{facchi2002quantum}, i.e., omitting the highly-oscillating terms for $g \gg \Omega$. Then we have a new simplified Hamiltonian as below,
\begin{equation}
\eqalign{
\hat{\mathcal{H}}_{eff} &= \frac{\Omega}{2}[\ket{\Psi_0}(\bra{\phi_1} - \bra{\phi_{14}})] + \Delta_1\ket{\Psi_0}\bra{\Psi_0} + \frac{\Delta_1}{2}\sum_{m=1}^{3}\ket{\Psi_m^{\pm}}\bra{\Psi_m^{\pm}} \cr
& + \frac{\Delta_1}{2}\sum_{m=1}^{3}(e^{i(\lambda_m^{+}-\lambda_m^{-})t}\ket{\Psi_m^{+}}\bra{\Psi_m^{-}} + {\rm H.c.}.
}
\label{eq:10}
\end{equation}
Despite the $\ket{\Psi_m^+}$ and $\ket{\Psi_m^-}$ which is decoupled to $\{\ket{\Psi_0},\ket{\phi_1},\ket{\phi_{14}}\}$,
the system can be described as a $\Lambda$-type three-level quantum system possessing an upper state $\ket{\Psi_0}$ and two lower states $\ket{\phi_1}$ and $\ket{\phi_{14}}$. 


\subsection{Effective model}

Starting from Eq.~(\ref{eq:10}), for large detuning condition $\Delta_1 \gg \Omega$, $\hat{\mathcal{H}}_{eff}$ could be further simplified as
\begin{equation}
\hat{\mathcal{H}}_{eff} = \frac{\Omega^{2}}{4\Delta_1}\ket{\phi_1}\bra{\phi_{14}} + {\rm H.c.},
\label{eq:11}
\end{equation}
which evolves from the initial state $\ket{\phi_1}$ to
\begin{equation}
\eqalign{
\ket{\psi(t)}&=\ket{f}_1\otimes[\cos{\frac{\Omega^2 t}{4\Delta_1}}\ket{g_Rg_L}_{23} - i\sin{\frac{\Omega^2 t}{4\Delta_1}}\ket{g_Lg_R}_{23}]\otimes\ket{0}.
}
\label{eq:12}
\end{equation}
The evolution on the Hilbert space corresponding to the original Hamiltonian Eq.~(\ref{eq:1}) is
\begin{equation}
\eqalign{
\ket{\psi(t)}&=\ket{f}_1\otimes[\cos{\frac{\Omega^2 t}{4\Delta_1}}\ket{g_Rg_L}_{23} - i\sin{\frac{\Omega^2 t}{4\Delta_1}}\ket{g_Lg_R}_{23}]\otimes\ket{000}_c\ket{00}_f.
}
\label{eq:13}
\end{equation}


\begin{figure}
\centering
\includegraphics[scale=0.6]{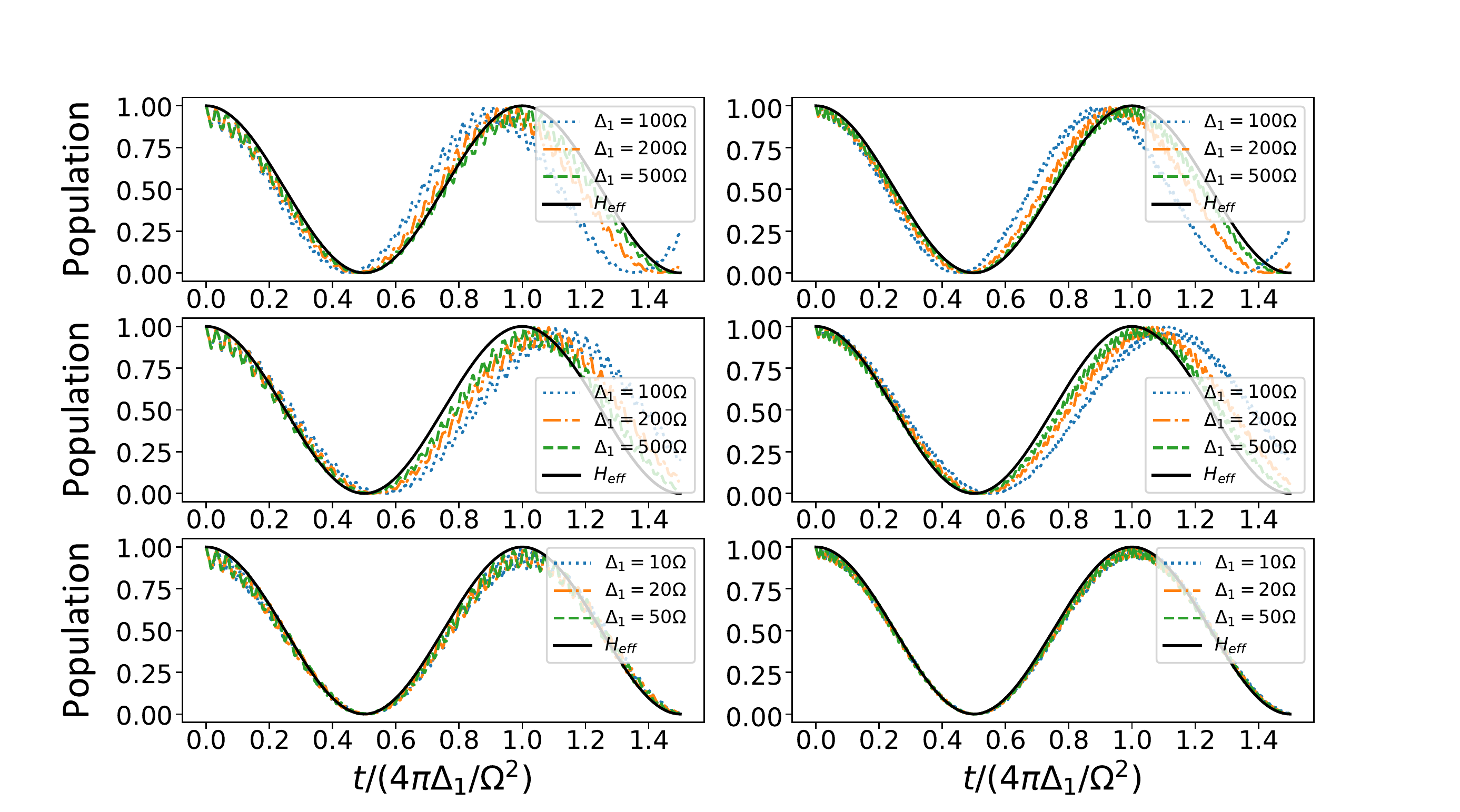}
\caption{Population of $\ket{\phi_0}$ in comparison of the original Hamiltonian with the effective Hamiltonian. (a1) $\nu = 5g = 25\omega$ and (b1)
$\nu = 5\sqrt{2}g = 50\Omega$, where other parameters are $\Delta_2 = \Delta_1 - /\sqrt{3}\nu$, $g_{c1}=\sqrt{6}g/2$ and
$g_{c2}=2\sqrt{3}g$. (a2)$\nu = 5g = 25\omega$ and (b2)$\nu = 5\sqrt{2}g = 50\Omega$, where other parameters are $\Delta_2 = \Delta_1 + \sqrt{3}\nu$,
$g_{c1}=\sqrt{6}g/2$ and $g_{c2}=2\sqrt{3}g$. (a3)$\nu = 5g = 25\omega$ and (b3)$\nu = 5\sqrt{2}g = 50\Omega$, where other parameters used are $\Delta_1 = \Delta_2$, $g_{c1}=-\sqrt{6}g/2$ and $g_{c2}=\sqrt{3}g$. }
\label{fig:5}
\end{figure}



\section{Application}
\subsection{QESD}

Now from the effective Hamiltonian $\hat{\mathcal{H}}_{eff}$ in Eq.~(\ref{eq:10}) with the initial state $\ket{\phi_0}$, we tune $\Delta_1$ and $\Omega$,
and the system evolves to
\begin{equation}
\eqalign{
\ket{\psi(t)}&=\ket{f}_1\otimes[\cos{\omega t}\ket{g_Rg_L}_{23} - i\sin{\omega t}\ket{g_Lg_R}_{23}]\otimes\ket{000}_c\ket{00}_f,
}
\label{eq:14}
\end{equation}
where $\omega$ is $\Omega^2/4\Delta_1$. The result clearly shows that throughout the evolution, atom 1 keeps staying in the state $\ket{f}_1$ and the bosonic mode
remains in vacuum state, whereas atom 2 and atom 3 turn to be entangled. The maximum entanglement occurs at $\tau=\pi/4\omega$ yielding the target state
$\ket{\psi_{tar}}=\frac{1}{\sqrt{2}}\ket{f}_1\otimes(\ket{g_Rg_L}_{23}-i\ket{g_Lg_R}_{23})\otimes\ket{000}_c\ket{00}_f$. If we have a $\pi/2$-phase operation on atom 2, the system will be a standard Bell
state $\ket{Bell}=\frac{1}{\sqrt{2}}\ket{f}_1\otimes(\ket{g_Rg_L}_{23}+\ket{g_Lg_R}_{23})\otimes\ket{000}_c\ket{00}_f$.


\subsection{QST}

Based on Eq.~(\ref{eq:13}), we may achieve the QST for arbitrary quantum states. For example, for an initial quantum state
\begin{equation}
\ket{\psi_0} = \ket{f}_1\otimes[\alpha\ket{g_R}_{2} + \beta\ket{g_L}_{2}]\otimes\ket{g_L}_3\otimes\ket{000}_c\ket{00}_f,
\label{eq:15}
\end{equation}
an evolution for $\omega t = \pi/2$ and then a $\pi/2$-phase operation on atom 3 could yield the state transfer from atom 2 to atom 3 as below,
\begin{equation}
\ket{\psi_{QST}} = \ket{f}_1\otimes \ket{g_L}_2\otimes[\alpha\ket{g_R}_{3} + \beta\ket{g_L}_{3}]\otimes\ket{000}_c\ket{00}_f.
\label{eq:16}
\end{equation}

\begin{figure}
\centering
\includegraphics[scale=0.5]{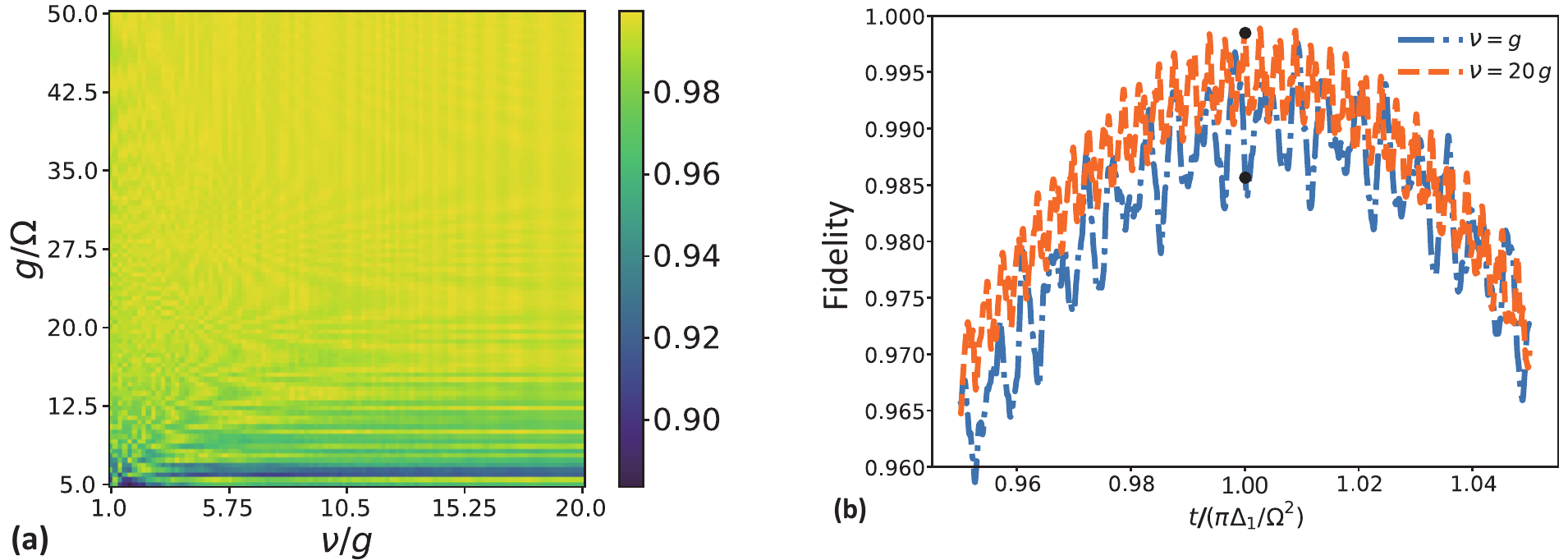}
\caption{(a)$F$ of the original Hamiltonian with respect to $g$ and $\nu$ at $\tau = \pi\Delta_1/\Omega^{2}$, where $\Delta_1=\Delta_2= 20\Omega$, $g_{c1}=-\sqrt{6}g/2$ and $g_{c2}=\sqrt{3}g$.(b) The effect on local evolution fluctuation due to different $\nu$ when $g = 20\Omega$. }
\label{fig:6}
\end{figure}

\section{Numerical simulation}

Since we have simplified the original Hamiltonian by a series of approximations, we have to justify the effective Hamiltonian after simplification. In order to make the simplified model hold, the relationship that needs to be satisfied are $\nu \gg g \gg \Omega$ and $\Delta_1 \gg \Omega$. Here we exemplify the QESD and check numerically the validity of those approximations by comparing the original Hamiltonian with the effective one. 

\subsection{Different parameter conditions}

In this subsection, we check three groups of parameter conditions, ${\delta_{+\sqrt{3}},\delta_{-\sqrt{3}},\delta_0}$, by comparing the results of time evolution of $\ket{\phi_0}$ calculated from the original and the effective Hamiltonians Eq.~(\ref{eq:10}).

\begin{enumerate}
\item $\delta_{+\sqrt{3}} = 0$
\label{item 1}

In this situation, the parameter relationship that needs to be satisfied is $ \Delta_1 - \Delta_2 = \sqrt{3}\nu \gg g \gg \Omega$ and $\Delta_1 \gg \Omega$. As plotted in Fig. \ref{fig:5}(a1,b1), $\Delta_1$ mainly decides the frequency of the evolution while $g$ and $\nu$ influence the local fluctuation. This situation needs a large $\Delta$ which means a long operation time $\tau = \pi\Delta/\Omega^2$ should be a long time.

\item $\delta_{-\sqrt{3}} = 0$
\label{item 2}

Nearly same as (\ref{item 1}), the parameter relationship that needs to be satisfied is $ \Delta_2 - \Delta_1 = \sqrt{3}\nu \gg g \gg \Omega$ and $\Delta_1 \gg \Omega$. As plotted in Fig. \ref{fig:5}(a2,b2), the result is really similar to (\ref{item 1}). That is the reason why the calculation of (\ref{item 2}) is similar to (\ref{item 1}).


\begin{figure}
\centering
\includegraphics[scale=0.6]{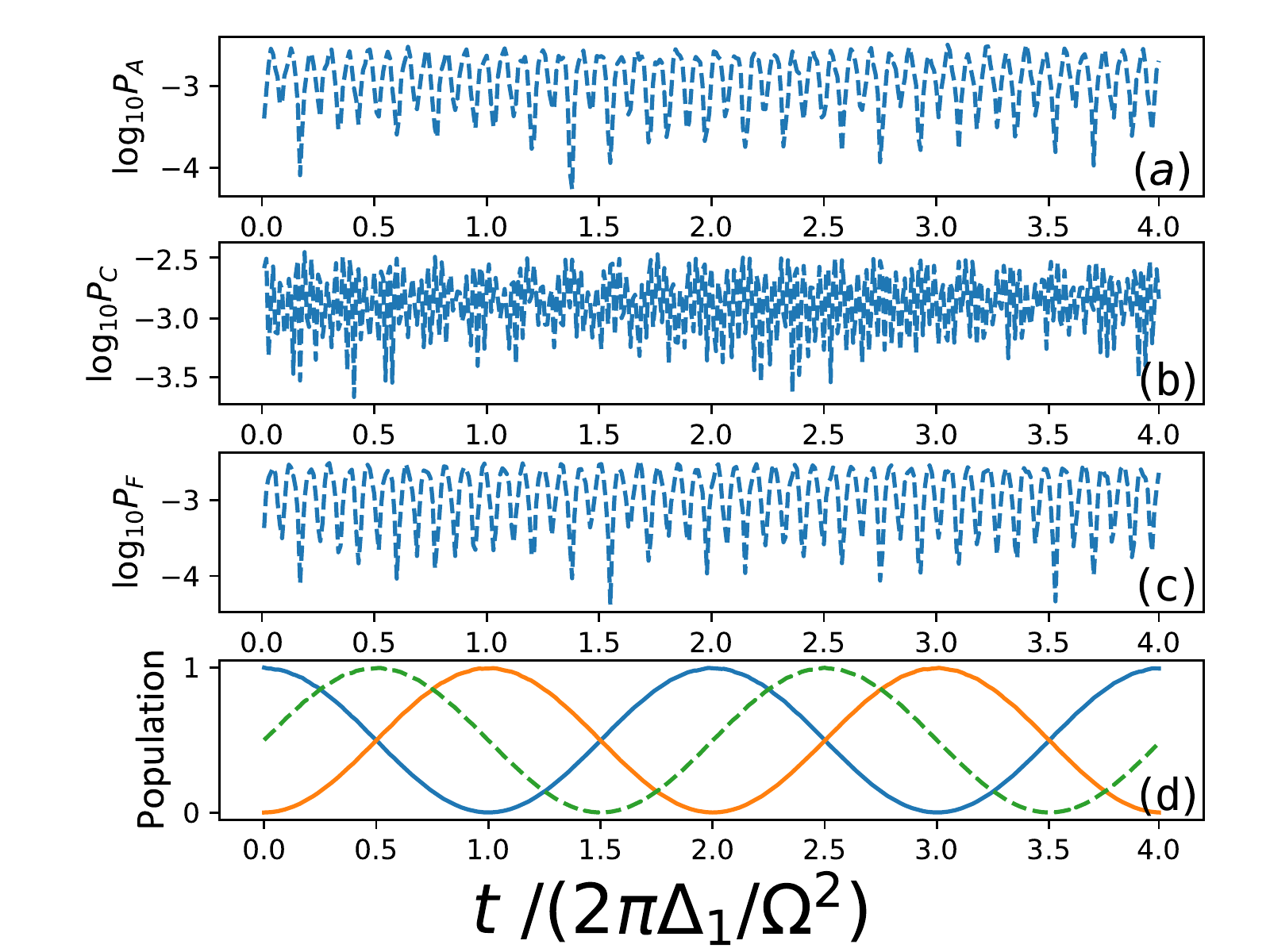}
\caption{Dynamics of the system, where (a) $\log_{10}(P_{A})$ with $P_{A}$ being the sum of the all atomic excited-state populations; (b) $\log_{10}(P_{C})$ with
$P_{C}$ being the sum of the populations
of the cavity's nonzero photon-number states; (c) $\log_{10}(P_{F})$ with $P_{F}$ being the sum of the populations of the fiber's nonzero photon-number states;  (d)
population in $\ket{g_{R}g_{L}}$ (blue solid), $\ket{g_{L}g_{R}}$ (brown solid), and $F$ (green dashed) which is the fidelity of creating the target state $|\psi_{tar}\rangle$. Parameters used here: $g=30\Omega$, $g_{c1}=-\sqrt{6}g/2$, $g_{c2}=\sqrt{3}g$, $\nu = 50\Omega$, $\Delta_1=\Delta_2=20\Omega$.}
\label{fig:7}
\end{figure}


\item $\delta_0 = 0$
\label{item 3}

In this situation, $\nu$ in independent from $\Delta_1$ and $\Delta_2$. The parameter relationship that needs to be satisfied is $\nu \gg g \gg \Omega$ and $\Delta_1 \gg \Omega$. As plotted in Fig. \ref{fig:5}(a3,b3), when $\Delta_1 = \Delta_2$, the results of the effective model and the actual model match very well. 

\end{enumerate}

From the fitting in Fig. \ref{fig:5}, we know that the frequency of the evolution in our scheme is mainly controlled by detuning $\Delta_1$ and $\Delta_2$. Meanwhile, local fluctuation is caused by hopping strength $\nu$ and coupling strength $g$. In consideration of the impact of operation time, we adopt the scheme in (\ref{item 3}) for further discussion. 

The fidelity of system is calculated by 
\begin{equation}
F = \bra{\psi_{tar}}\hat{\rho}(\tau)\ket{\psi_{tar}}
\label{eq:17}
\end{equation}
where $\ket{\psi_{tar}}$ is target state that we want to implement, $\hat{\rho}(\tau)$ denotes the density operator of this system at operation time $\tau$. Then, we check the validity of quantum Zeno condition $g \gg \Omega$ and hopping strength $\nu$. The result in Fig. \ref{fig:6}(a) reveals $g$ mainly affects the fidelity of the system. When $g \ge 20\Omega$, the effective model also has a high fidelity when the condition $\nu \gg g$ is not fully satisfied. However the frequency of fluctuation will decrease when $\nu$ takes a small value in Fig. \ref{fig:6}(b). Experimentally, it is relatively difficult to achieve high-strength fiber coupling $\nu$. This numerical result shows that for obtaining relatively high fidelity, the model allows fiber coupling satisfy $\nu \sim g$ as long as $g \gg \Omega$.


\subsection{Validity of the virtual photon}

One of the advantages of our scheme is the achievement of entanglement and state transfer between the distant nodes via virtual photon effects. As presented in Fig. \ref{fig:7}, we justify
this virtual photon condition numerically, in which the approximation can be found to work nearly perfect. The population in excited states, cavities and fibers are all less than 0.01 and this illustrates that the system is robust. In next section, we will discuss the robustness of system in details.


\subsection{Robustness against decoherence}

Taking decoherence into consideration, we check the evolution of the whole system by Lindblad master equation,
\begin{equation}
\eqalign{
\dot{\rho} &= i[\rho,\hat{H}_{I}] +
\frac{1}{2}\sum_{k=1}^{3}\sum_{j=L,R}[2\mathcal{L}_{k,j}\rho\mathcal{L}^{\dag}_{k,j}-(\mathcal{L}_{k,j}\mathcal{L}^{\dag}_{k,j}\rho +
\rho\mathcal{L}^{\dag}_{k,j}\mathcal{L}_{k,j})] \cr
& + \frac{1}{2}\sum_{m=1}^{3}\sum_{j=L,R}[2\mathcal{L}_{m,j}\rho\mathcal{L}^{\dag}_{m,j}-(\mathcal{L}_{m,j}\mathcal{L}^{\dag}_{m,j}\rho +
\rho\mathcal{L}^{\dag}_{m,j}\mathcal{L}_{m,j})] \cr
& + \frac{1}{2}\sum_{n=1}^{2}\sum_{j=L,R}[2\mathcal{L}_{n,j}\rho\mathcal{L}^{\dag}_{n,j}-(\mathcal{L}_{n,j}\mathcal{L}^{\dag}_{n,j}\rho +
\rho\mathcal{L}^{\dag}_{n,j}\mathcal{L}_{n,j})] \cr
& + \frac{1}{2}[2\mathcal{L}_{0}\rho\mathcal{L}^{\dag}_{0}-(\mathcal{L}_{0}\mathcal{L}^{\dag}_{0}\rho + \rho\mathcal{L}^{\dag}_{0}\mathcal{L}_{0})],
}
\label{eq:18}
\end{equation}
where $\mathcal{L}_{0}=\sqrt{\gamma_{1}}\ket{f}_{1}\bra{e}$, $\mathcal{L}_{k,j}=\sqrt{\gamma_{k,j}}\ket{g_{j}}_{k}\bra{e}$,
$\mathcal{L}_{m,j}=\sqrt{\kappa_{m,j}}\hat{a}_m$ and
$\mathcal{L}_{n,j}=\sqrt{\kappa_{n,j}}\hat{b}_{n}$ describe various docoherece effects in the system. To simplify our treatment, we assume
$\gamma_{1}=\gamma_{1,j}=\gamma/3$,
$\gamma_{2,j}=\gamma_{3,j}=\gamma/2$ and $\kappa_{m,j}=\kappa_{n,j}=\kappa/2$ with the spontaneous emission rate $\gamma$ of each atom and the photon leakage rate
$\kappa$ of the cavity or fiber.

\begin{figure}
\centering
\includegraphics[scale=0.6]{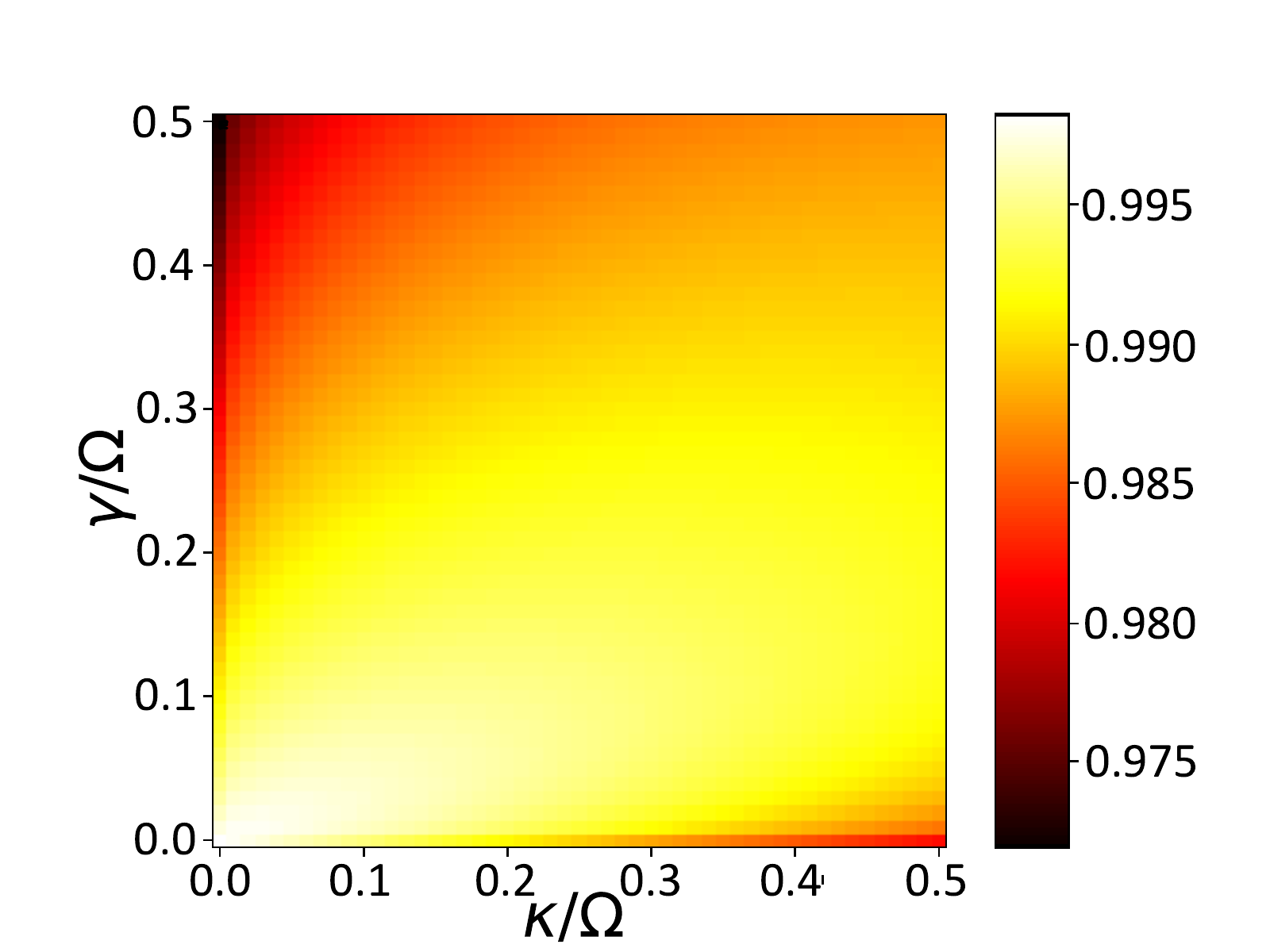}
\caption{Fidelity of the target state at $\tau=\pi\Delta_1/\Omega^{2}$ under different values of dissipation. Parameters used here: $g=30\Omega$, $g_{c1}=-\sqrt{6}g/2$, $g_{c2}=\sqrt{3}g$, $\nu = 50\Omega$,
$\Delta_1=\Delta_2=20\Omega$}
\label{fig:8}
\end{figure}

We plot in Fig. \ref{fig:8} the fidelity, with respect to the ideal case, as functions of $\gamma/\Omega$ and $\kappa/\Omega$ at evolving time $t =
\pi\Delta_1/\Omega^{2}$. This scheme is very robust against decoherence induced by
atomic spontaneous emissions and photonic leakages from the cavity-fiber system. We see from Fig. \ref{fig:8} that, when $\gamma$ and $\kappa$ are around $0.5\Omega$, the fidelity at $t$ can be
still beyond 0.94. The atomic spontaneous emissions rate $\gamma$ influences the system more than other decaying factors.

\section{Experimental feasibility}

\begin{figure}[htbp]
\centering
\includegraphics[scale = 0.6]{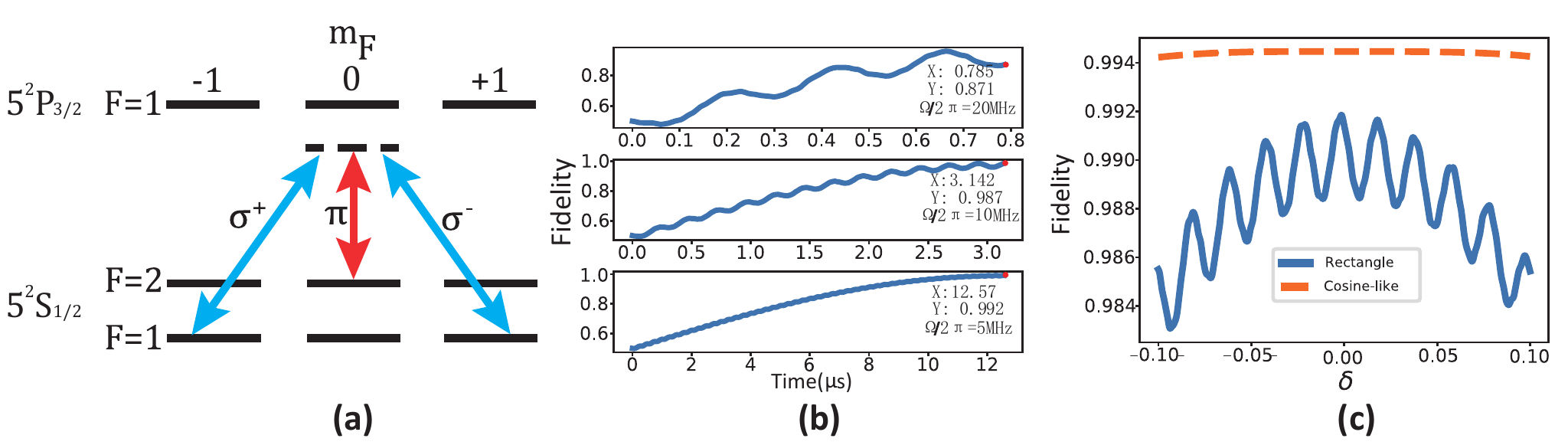}
\caption{(a) Energy levels and related transitions in $^{87}$Rb atoms, where $\sigma^{+}$, $\sigma^{-}$ and $\pi$ denote the left-circular, right-circular and linear
polarizations, respectively; (b) Time evolution of the fidelity for creating the target state with different values of $\Omega$, where $g_{n,j}/2\pi = 300$ MHz, $\gamma/2\pi = 3$ MHz, $\kappa_c/2\pi = 1.5$ MHz, $\kappa_f/2\pi = 152$ kHz, $\nu/2\pi = 300$ MHz, and $\Delta_1(\Delta_2)/2\pi=100$ MHz. X and Y donate the operation time and fidelity at red point.(c)
Fidelity with respect to the deviation from the ideal operation time.}
\label{fig:9}
\end{figure}

The system under consideration could be realized in cold alkali-metal atoms, such as $^{135}$Cs or $^{87}$Rb \cite{wilk2007single,lettner2011remote,AlkaliData}, as
considered in Fig. \ref{fig:9}(a). Based on recent experimental reports employing high-Q cavities and strong atom-cavity coupling
\cite{spillane2003ideality,vernooy1998high,armani2003ultra,buck2003optimal,spillane2005ultrahigh,barclay2006pe}, we may choose the parameters as $g_{n,j}/2\pi\sim
300$ MHz,
$\gamma/2\pi\sim 7.5$ MHz and $\kappa_c/2\pi\sim 1.5$ MHz. The fiber decay rate can be set as $\kappa_f/2\pi\sim 152$ kHz \cite{gordon2004short}.
Using these parameters, we simulate our scheme with different values of $\Omega$, as shown in Fig. \ref{fig:9}(b), where the fidelity is about $99.2\%$ after the system evolves for $12.5 \mu$s under $\Omega/2\pi = 5$ MHz.


In order to minimize the influence from the experimental imperfection, we try to accelerate the implementation as
discussed above. From Eq.~(\ref{eq:10}) we know the final fidelity depending on $\int{\Omega^2}dt$. As such, we choose a cosine-like function,
\begin{equation}
\Omega(t)=\Omega_m[\cos{(\frac{2\pi t}{T^{\prime}}-\pi)}+1]/2,
\label{eq:19}
\end{equation}
where $\Omega_m$ is the maximum amplitude. To satisfy $\int_{0}^{T^\prime}\Omega(t)^2dt=\int_{0}^{T}\Omega^2dt$, we obtain $3\Omega_m^2T^\prime=8\Omega^2T$, implying that a larger $\Omega_m$,
could effectively accelerate entanglement generation and QST. Fig. \ref{fig:9}(c) indicates that the laser pulse with cosine-like function works much better than the usual rectangular form.

\section{Conclusion}

To summarize, we have proposed a practical scheme to achieve QESD and QST in an atom-cavity-fiber model, which could work for future quantum network. The three
favorable features, i.e., the auxiliary atom under laser driving always in the ground state, no excitation for every atom and field mode throughout implementation, and no actual
operation performed on the atoms for entanglement, make our scheme experimentally feasible with current laboratory techniques and robust to experimental imperfection.
In this context, we argue that our scheme is easily extended to multi-atom case with each cavity confining
N atoms, for which the coupling strength could become larger with more atoms involved and thus less operation time is required. We argue that our scheme would be
helpful for exploiting quantum network connected by optical fibers or even in wireless way. Finally, we suggest to choose $\Delta_1 = \Delta_2$, under which the laser action time can be decreased greatly and $\nu$ is independent of $\Delta_1$ and $\Delta_2$. In addition, the value of $\nu$ can take $\nu \sim g$ when $g \gg \Omega$.


\section*{Acknowledgments}

We thank Qutip \cite{johansson2012qutip} for its open source library of python for our numerical simulations. This work was supported by National Key Research
and Development Program of China under Grant No. 2017YFA0304503 and by National Natural Science Foundation
of China under Grants No. 11804308, No. 11835011, No. 11804375, No. 11734018 and No. 11674360.


\section*{References}
\bibliographystyle{iopart-num}
\bibliography{REV}

\appendix
\section{}

Choosing different detuning $\delta_n$ we could reduce Eq.~(\ref{eq:4}) into following effective Hamiltonians,
\begin{enumerate}
\item $\delta_{+\sqrt{3}}=0$
\begin{equation}
\eqalign{
\hat{H}_{eff1} &= \Omega\ket{e}_1\bra{f}+\sum_{j=L,R}(\frac{g_{1,j}}{\sqrt{3}}\hat{c}_{+\sqrt{3},j}\ket{e}_1\bra{g_{j}} \cr
& + \sum_{k=2}^{3}\frac{g_{k,j}}{2\sqrt{3}}\hat{c}_{+\sqrt{3},j}\ket{e}_{k}\bra{g_{j}}) + {\rm H.c.}] + \Delta_1\sum_{j=1}^{3}\ket{e}_j\bra{e},
}
\label{eq:Heff1}
\end{equation}
where $\delta_{-\sqrt{3}}, \delta_{+}, \delta_{-}, \delta_{0} \gg g$. In large detuning limit, we have conditions $(\sqrt{3}+1)\nu \gg g$, $(\sqrt{3}-1)\nu \gg g$ and $\sqrt{3}\nu \gg g$.
\item $\delta_{-\sqrt{3}}=0$
\begin{equation}
\eqalign{
\hat{H}_{eff2}& = [\Omega\ket{e}_1\bra{f}+\sum_{j=L,R}(\frac{g_{1,j}}{\sqrt{3}}\hat{c}_{-\sqrt{3},j}\ket{e}_1\bra{g_{j}} \cr
& + \sum_{k=2}^{3}\frac{g_{k,j}}{2\sqrt{3}}\hat{c}_{-\sqrt{3},j}\ket{e}_{k}\bra{g_{j}}) + {\rm H.c.}] + \Delta_1\sum_{j=1}^{3}\ket{e}_j\bra{e},
}
\label{eq:Heff2}
\end{equation}
where $\delta_{+\sqrt{3}}, \delta_{+}, \delta_{-}, \delta_{0} \gg g$. In large detuning limit, we have conditions $(\sqrt{3}+1)\nu \gg g$, $(\sqrt{3}-1)\nu \gg g$ and $\sqrt{3}\nu \gg g$.
\item $\delta_0=0$
\begin{equation}
\eqalign{
\hat{H}_{eff3}& = [\Omega\ket{e}_1\bra{f}+\sum_{j=L,R}(-\frac{g_{1,j}}{\sqrt{3}}\hat{c}_{0,j}\ket{e}_1\bra{g_{j}} \cr
& + \sum_{k=2}^{3}\frac{g_{k,j}}{\sqrt{3}}\hat{c}_{0,j}\ket{e}_{k}\bra{g_{j}}) + {\rm H.c.}] + \Delta_1\sum_{j=1}^{3}\ket{e}_j\bra{e}.
}
\label{eq:Heff3}
\end{equation}
where $\delta_{-\sqrt{3}}, \delta_{+\sqrt{3}}, \delta_{+}, \delta_{-}, \gg g$. In large detuning limit, we have conditions $\sqrt{3}\nu \gg g$ and $\nu \gg g$.

\item $\delta_+=0$
\begin{equation}
\eqalign{
\hat{H}_{eff4}& = [\Omega\ket{e}_1\bra{f}+ \sum_{k=2}^{3}\frac{g_{k,j}}{2}\hat{c}_{+,j}\ket{e}_{k}\bra{g_{j}}) + {\rm H.c.}] + \Delta_1\sum_{j=1}^{3}\ket{e}_j\bra{e}.
}
\label{eq:Heff4}
\end{equation}
where $\delta_{+\sqrt{3}}, \delta_{-\sqrt{3}},\delta_{0}, \delta_{-} \gg g$. In large detuning limit, we have conditions $(\sqrt{3}+1)\nu \gg g$, $(\sqrt{3}-1)\nu \gg g$ and $\nu \gg g$.

\item $\delta_-=0$
\begin{equation}
\eqalign{
\hat{H}_{eff5}& = [\Omega\ket{e}_1\bra{f}+ \sum_{k=2}^{3}\frac{g_{k,j}}{2}\hat{c}_{-,j}\ket{e}_{k}\bra{g_{j}}) + {\rm H.c.}] + \Delta_1\sum_{j=1}^{3}\ket{e}_j\bra{e}.
}
\label{eq:Heff5}
\end{equation}
where $\delta_{+\sqrt{3}}, \delta_{-\sqrt{3}}, \delta_{0}, \delta_{+} \gg g$. In large detuning limit, we have conditions $(\sqrt{3}+1)\nu \gg g$, $(\sqrt{3}-1)\nu \gg g$ and $\nu \gg g$.

\end{enumerate}
So to summarize, $\nu \gg g$ should be satisfied in Eq.~(\ref{eq:5}).

In order to describe $\hat{H}_{eff}$ in a single-exciton space, we introduce a set of bases as below,
\begin{equation}
\eqalign{
\ket{\phi_{1}}=\ket{f}_1\ket{g_Rg_L}_{23}\ket{0},&\ket{\phi_{2}}=\ket{e}_1\ket{g_Rg_L}_{23}\ket{0}, \cr
\ket{\phi_{3}}=\ket{g_R}_1\ket{g_Rg_L}_{23}\ket{R},&\ket{\phi_{4}}=\ket{g_L}_1\ket{g_Rg_L}_{23}\ket{L}, \cr
\ket{\phi_{5}}=\ket{g_R}_1\ket{eg_L}_{23}\ket{0},&\ket{\phi_{6}}=\ket{g_L}_1\ket{g_Re}_{23}\ket{0}, \cr
\ket{\phi_{7}}=\ket{g_R}_1\ket{g_Lg_L}_{23}\ket{L},&\ket{\phi_{8}}=\ket{g_L}_1\ket{g_Rg_R}_{23}\ket{R}, \cr
\ket{\phi_{9}}=\ket{g_R}_1\ket{g_Le}_{23}\ket{0},&\ket{\phi_{10}}=\ket{g_L}_1\ket{eg_R}_{23}\ket{0}, \cr
\ket{\phi_{11}}=\ket{g_R}_1\ket{g_Lg_R}_{23}\ket{R},&\ket{\phi_{12}}=\ket{g_L}_1\ket{g_Lg_R}_{23}\ket{L}, \cr
\ket{\phi_{13}}=\ket{e}_1\ket{g_Lg_R}_{23}\ket{0},&\ket{\phi_{14}}=\ket{f}_1\ket{g_Lg_R}_{23}\ket{0}.
}
\end{equation}

Then we reach Eq.~(\ref{eq:6}) in the main text.


\end{document}